\documentclass[letterpaper,twocolumn,10pt]{article}
\usepackage{usenix2019_v3}

\usepackage{tikz}
\usepackage{amsmath}
\usetikzlibrary{shapes, arrows}

\usepackage{filecontents}

\tikzstyle{block} = [rectangle, draw, 
    text width=8em, text centered, rounded corners, minimum height=2em]
\tikzstyle{line} = [draw, -latex']

\DeclareUnicodeCharacter{0308}{.}
\begin{document}
%-------------------------------------------------------------------------------

%don't want date printed
\date{}

\title{\Large \bf No RISC No Reward:\\
  Return-Oriented Programming on RISC-V}

%for single author (just remove % characters)
\author{
{\rm Garrett Gu}\\
UT Austin
\and
{\rm Hovav Shacham}\\
UT Austin
% copy the following lines to add more authors
% \and
% {\rm Name}\\
%Name Institution
} % end author

\maketitle

\begin{abstract}
RISC-V is an open-source hardware ISA based on the RISC design principles, and has been the subject of some novel ROP mitigation technique proposals due to its open-source nature. However, very little work has actually evaluated whether such an attack is feasible assuming a typical RISC-V implementation. We show that RISC-V ROP can be used to perform Turing complete calculation and arbitrary function calls by leveraging gadgets found in a version of the GNU libc library. Using techniques such as self-modifying ROP chains and algorithmic ROP chain generation, we demonstrate the power of RISC-V ROP by creating a compiler that converts code of arbitrary complexity written in a popular Turing-complete language into RISC-V ROP chains. 
\end{abstract}

\section{Introduction}
\label{section:intro}

RISC-V is an open RISC architecture developed at Berkeley and backed
by an international consortium.  Like other modern architectures,
RISC-V has hardware support for data execution prevention (DEP).

Return-oriented programming, or ROP, is a systematization and
generalization of code-reuse attack techniques like
return-into-libc~\cite{solar-designer} and the borrowed-chunks
technique~\cite{krahmer}.  With ROP, short instruction sequences, each
ending in a return instruction, are combined into building blocks
called gadgets.  An attacker who controls the stack can induce any
desired functionality by sequencing gadgets, even in the presence of
DEP.  ROP was originally described for the x86~\cite{ROP1}, and
subsequently extended to other architectures such as
SPARC~\cite{sparcrop}, ARM~\cite{armrop}, and PowerPC~\cite{rop-ppc}.

The RISC-V instruction set was designed for extensibility, with a
substantial portion of the opcode space left
unassigned~\cite{riscv-design}.  Researchers have taken advantage of
RISC-V's extensibility to propose hardware extensions that make
code-reuse attacks harder to mount (see, e.g.,
\cite{fingeret-thesis,p-taxi,zipper-stack,fixer}).

There's just one problem: No one seems to have shown that general,
Turing-complete ROP is possible on RISC-V.\footnote{In independent,
  simultaneous work to appear at AsiaCCS in October, Jaloyan
  et~al.~\cite{riscv-rop-asiaccs} describe more general ROP techniques
  for RISC-V.  We were unable to find a preprint of the Jaloyan
  et~al.\ paper as of July 1, 2020.}

In this short note, we close this gap by showing that ROP is, indeed,
possible on RISC-V.  We demonstrate how arithmetic, memory reads and
writes, conditional branching, and function invocation can be
performed solely through ROP gadgets found in the GNU libc library.
To work around limitations in the available instruction sequences, we
introduce techniques for self-modifying ROP chains that may be of
independent interest.  As is now traditional (see, e.g.,
\cite{run-dma}), we show that our gadget set is Turing complete by
designing a compiler that accepts a Brainfuck program of arbitrary
complexity and generates an equivalent RISC-V ROP chain.

\subsection{Experimental Platform}
\label{section:platform}

We analyze glibc-2.30.9000-29.fc32.riscv64, runninng on
Fedora-Developer-Rawhide-20200108.n.0 booting on the virt machine in
QEMU. Since we do not leverage any features of emulated hardware
outside of documented ISA, our ROP chains and gadgets should also work
on equivalent hardware.

We target the RV64GC variant of the RISC-V ISA, which includes the
64-bit base integer instruction set as well as the MAFDC extensions:
integer multiplication/division, atomic instructions, single- and
double-precision floating point, and compressed instructions.

The C extension would be interesting for analyzing ROP potential
because it allows for some degree of variable-length
instructions. Rather than all instructions being 4 bytes long, the C
extension allows certain commonly-used instructions to only take up 2
bytes in an effort to reduce code size.

Unlike ARM Thumb mode, the RISC-V C extension does not require a mode
switch, and compressed instructions can coexist with uncompressed
instructions in the same instruction stream. We will analyze the
implications of this optional but very common extension on ROP gadget
availability.

\section{Complexity of RISC-V ROP gadgets}
\label{section:gadgetcomplexity}
RISC-V makes use of a link register, like similar RISC architectures such as PowerPC, ARM, and SPARC. In RISC-V this is labeled the \texttt{ra} register. The purpose of this register is to optimize calls to leaf subroutines since the return address need not be pushed or popped on the stack. The implication of this is that in RISC ROP exploitation is mostly limited to non-leaf function epilogues, and RISC-V ROP is no exception.

For our purposes, we define a \textbf{chainable ROP gadget} as follows: an instruction sequence that:
\begin{enumerate}
  \item Loads a value from \texttt{a(sp)} into \texttt{ra} where \texttt{a} is some positive immediate value divisible by 8
  \item Adds an immediate value \texttt{b} to \texttt{sp} where \texttt{b > a} and \texttt{b} is divisible by 16 (due to stack-alignment requirements)
  \item Ends in a \texttt{ret} (equivalent to \texttt{jr ra})
\end{enumerate}
In addition to the above steps needed to maintain the ROP chain, a good ROP gadget will contain a few extra instructions that perform useful work. Note that these requirements are fairly restrictive, and exclude gadgets that, for example, pop and jump to a register other than the \texttt{ra} register. For our purposes, we will focus on gadgets that fulfill these requirements (and thus look like function epilogues) but examining other classes of gadgets is a promising future research direction. 

\subsection{A minimal ROP gadget}
\label{section:nopgadget}
As an example, below is a gadget that does no work except maintining the ROP chain. This is typically called a NOP gadget.
\begin{verbatim}
  0x0000000000097a68 :
      c.ldsp ra, 8(sp)
      c.addi sp, 0x10
      c.jr ra
\end{verbatim}

Note that in x86, the same three actions are performed by a single 1-byte instruction (\texttt{ret}). In SPARC, a similar gadget only requires a \texttt{ret} and \texttt{restore} to slide the register window. In ARM, a similar gadget looks like a \texttt{LDMFD} followed by a \texttt{RET}.

The fact that the simplest ROP gadget in RISC-V requires more instructions than in other architectures has the following implications:
\begin{enumerate}
  \item When a chainable ROP gadget is found, it is very likely that the gadget formed part of a legitimate function epilogue. ROP gadgets that consist entirely of unintended instructions would be exceedingly rare.
  \item The three instructions may not always be \textit{contiguous}; in other words, part of the work performed by the gadget may be located in between these three instructions. When this happens, the return-oriented programmer is unable to "trim" the work by jumping into the middle of the work like on other architectures. Even in ARM, the \texttt{LDMFD} and the \texttt{RET} are typically contiguous. Similarly, in SPARC, \texttt{ret} and \texttt{restore} are almost always contiguous. The implication of RISC-V's departure from this norm is that RISC-V ROP chains sometimes must account for a larger number of undesirable gadget side effects.
\end{enumerate}

Take for example the following POP gadget:
\begin{verbatim}
  0x000000000006a5e8 : 
      c.ldsp ra, 0x28(sp)
      c.ldsp s0, 0x20(sp)
      c.ldsp a0, 0(sp)
      c.ldsp a1, 8(sp)
      c.ldsp s1, 0x18(sp)
      c.ldsp s2, 0x10(sp)
      c.addi16sp sp, 0x30
      c.jr ra
\end{verbatim}
If the Return-Oriented Programmer would like to use this gadget to pop \textit{only} \texttt{s2} (maybe a0 contains a runtime-calculated value she would not like to overwrite), she is unable to do so because jumping directly to the \texttt{c.ldsp s2, 0x10(sp)} instruction would skip the load into \texttt{ra}, breaking the ROP chain and causing an infinite loop. 

A similar issue presents itself with ARM through \texttt{LDMFD} instructions that pop a large number of registers, and with SPARC through the \texttt{restore} instruction. Note however, that in RISC-V, the instructions sandwiched between \texttt{c.ldsp ra, 0x28(sp)} and \texttt{c.addi16sp, sp, 0x30} are often not only pop instructions and thus can cause traps, undefined behavior, and undesirable memory corruption.

\subsection{Preconditions}
\label{section:preconditions}
The ideal way to avoid undesirable side effects is to entirely avoid using gadgets that cause them. However this is not always feasible, and sometimes more careful and deliberate treatment of side effects is needed.

Take for example the following readMEM gadget: 
\begin{verbatim}
  0x00000000000d3230 : 
      c.ld a0, 8(a0)
      c.add a0, a5
      c.ldsp a4, 0x28(sp)
      c.ld a5, 0(s0)
      bne a4, a5, 0x1e
      c.ldsp ra, 0x38(sp)
      c.ldsp s0, 0x30(sp)
      c.addi16sp sp, 0x40
      c.jr ra
\end{verbatim}

This gadget is extremely valuable because it is a very rare readMEM gadget that reads memory pointed at by a register that is easily popped, incremented, and decremented through other gadgets, and does not perform a conditional branch depending on the read value. However, it is prone to the following side effects:

\begin{enumerate}
  \item The initial value of \texttt{a5} is added to the read value of \texttt{a0}.
  \item The value of \texttt{a5} is read from \texttt{(s0)}.
  \item The value of \texttt{a4} is popped and if the popped \texttt{a4} is not equal to the read \texttt{a5} then a branch occurs. 
\end{enumerate}

Ideally, we want the initial value of \texttt{a5} to be 0 so that the first side effect is avoided entirely, and we want \texttt{(s0)} to not trap and contain some constant known value so that we can make the popped \texttt{a4} equal to the value, avoiding the branch. If we maintain these preconditions prior to invoking this gadget (for example, if we found and used a pop gadget for \texttt{s0}, \texttt{a4}, and \texttt{a5}), then we can prevent or account for these side effects. It turns out that these specific preconditions can be achieved using two other gadgets. 

In terms of crafting ROP chains, we can combine the gadgets used to fulfill preconditions and the gadget requiring the preconditions into a single logical unit to make programming large ROP chains easier. 

\section{Self-modifying ROP}
\label{section:selfmodifyingrop}
In order to mitigate the relative sparsity of side-effect-free, clean gadgets in RISC-V ROP, we propose a trick that improves the capability of these gadgets.

\subsection{Saving registers}
\label{section:registersaving}
For example, one issue coming from the previous readMEM gadget is the fact that the address in \texttt{a0} was overwritten with the value from memory. If we would like to keep the original value for later use, we could either find a way to move the value to some other register which is not cobbled by the readMEM gadget sequence, or we could write the value to some scratch space in memory so that a later readMEM can read it back. 

We propose a technique called "self-modifying ROP" that simplifies the second approach. A similar approach was used in Sigreturn-Oriented Programming in x86, but our approach goes a bit deeper and applies to "vanilla" ROP chains instead. \cite{srop}

Rather than using a readMEM gadget to restore the value into the register, we use a POP gadget, and the POP gadget frame itself is the destination we write the original value to. The closest analogue to this technique comes from SPARC ROP, where a value is written to a future gadget's register window in order to set the value when the window is \texttt{restore}d.

In this case, we have the option to use self-modifying ROP rather than writing to scratch space; we avoid having to allocate scratch space and make our gadget sequence more self-contained as a benefit. As we will show later, self-modifyng ROP is much more powerful and can also be used to circumvent several other limitations of gadgets. 

\subsection{Manufacturing a MOV gadget}
\label{section:selfmodifiedpops}
One problem we ran into while evaluating RISC-V Turing completeness was needing to write to a runtime-calculated memory address. Since it was easiest to find arithmetic gadgets on \texttt{a0}, we wanted to find a gadget that wrote to \texttt{(a0)} or a constant offset away from \texttt{(a0)}. However, we were only able to find nice gadgets with minimal side effects that write to memory locations parameterized by \texttt{s0}, \texttt{s1}, \texttt{a3}, \texttt{a4}, and a few other registers. However, we couldn't find nice MOV gadgets from \texttt{a0} to one of these gadgets with manageable side effects. In the case of \texttt{s0} and \texttt{s1}, this makes sense; these registers are designated as callee-saved registers, so any function that modifies these registers must restore the original value by the time it returns. 

Luckily, the fact that \texttt{s0} and \texttt{s1} are callee-saved registers means that POP gadgets are easily available for them. Thus, these writeMEM gadgets can be used to write to locations which are hardcoded into the ROP chain. Further, we can use the same self-modifying ROP technique to write the value of \texttt{a0} to a future \texttt{s0} POP gadget, in effect "manufacturing" a MOV gadget from \texttt{a0} to \texttt{s0}. This allows our writeMEM gadget to write to an arbitrary calculated address.

\begin{figure}[]
  \centering
  \begin{tikzpicture}
    \node [block] (begin) {begin};
    \node [block, below of=begin] (writea0) {write \texttt{a0}};
    \node [block, below of=writea0, node distance=1.1cm] (cobblea0) {gadget sequence that cobbles \texttt{a0}};
    \node [block, below of=cobblea0, node distance=1.6cm] (restorea0) {
      \begin{tikzpicture}
        \node [text width=6em] (poptext) {pop \texttt{a0}};
        \node [text width=6em, draw, below of=poptext, node distance=0.8cm]   (innernode) {popped value};
      \end{tikzpicture}
    };
    \node [right of=cobblea0, text width=6em, node distance=3cm] (writetarget) {write target};
    \path [line] (begin) -- (writea0);
    \path [line] (writea0) -- (cobblea0);
    \path [line] (cobblea0) -- (restorea0);
    \draw (writea0) -| (writetarget);
    \path [line] (writetarget) |- (1.2, -4.1);
  \end{tikzpicture}
  \caption{A simple example of self-modifying ROP. Here self-modifying ROP is used to restore the previous value of \texttt{a0} after it has been cobbled by some gadget sequence. Boxes indicate gadget frames.}
  \label{figure:write_to_pop}
\end{figure}
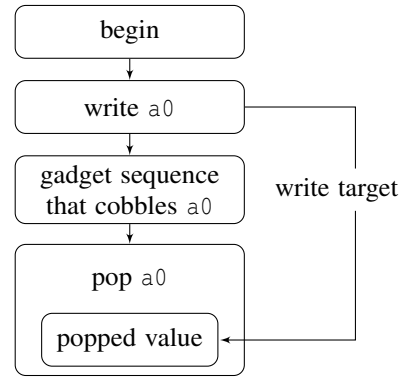

\subsection{Calling functions with arguments}
\label{section:functioncalls}
In order to call libc functions from the Return-Oriented Program, we found some POP gadgets and readMEM gadgets that can be used to place an arbitrary set of values into registers \texttt{a0}-\texttt{a5}. We also found a gadget that invokes \texttt{jalr a5}, which sets the return address and calls the function pointed to by \texttt{a5}. Using these gadgets, we can call any libc function with up to four parameters, all of which can be calculated at runtime using self-modifying ROP techniques. Since the found \texttt{jalr a5} gadget does not cobble \texttt{a0}, we can also retrieve the return value of a libc function.

A complication arises when the function would like to allocate stack space and write to it, corrupting the gadgets that come before the \texttt{jalr a5} gadget. This is not a problem when the gadgets execute in a linear fashion, but when we have loops and conditional branches (as we do), this corruption is an issue. We solve this issue in mostly the same way as SPARC ROP, by putting the \texttt{jalr a5} gadget frame immediately after a "safe" buffer zone that the function call is free to modify. We then invoke the \texttt{jalr a5} gadget using an unconditional branch. 

However, there is another complication. Our unconditional branch (part of glibc's \texttt{longjmp} function) reads the next gadget's \texttt{ra} and \texttt{sp} from \texttt{(a0)} and \texttt{0x68(a0)}, respectively. While this is not usually a problem, \texttt{a0} is also where the first parameter to a function is stored. Thus, if we were to directly branch to the \texttt{jalr a5} gadget, we would not be able to pass in an arbitrary value as the first parameter. We can solve this by dynamically creating a \texttt{a0} POP gadget frame located before the \texttt{jalr a5} gadget frame in memory, then branching to the \texttt{a0} POP gadget instead. Note that we have to create the \textit{entire} gadget frame including the popped \texttt{ra}, rather than simply write the popped value for \texttt{a0} like in the previous instances of self-modifying ROP. This is because the \texttt{a0} POP gadget frame itself is located within the "safe" zone of the function call and thus may be overwritten.

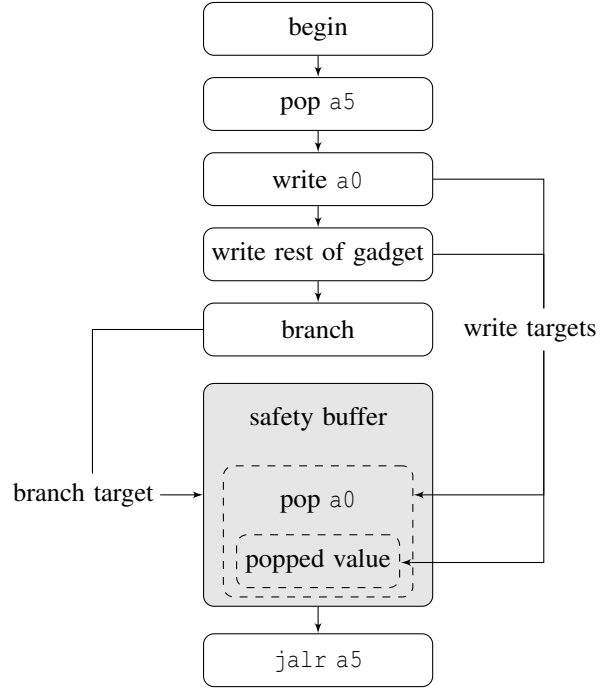
\begin{figure}[tb]
  \centering
  \begin{tikzpicture}
    \node [block] (begin) {begin};
    \node [block, below of=begin] (popa5) {pop \texttt{a5}};
    \node [block, below of=popa5] (writea0) {write \texttt{a0}};
    \node [block, below of=writea0] (writegadget) {write rest of gadget};
    \node [block, below of=writegadget] (branch) {branch};
    \node [block, below of=branch, node distance=2.2cm, fill=gray!20] (safetybuffer) {
      \begin{tikzpicture}
        \node [text width=6em] (safetybuffertext) {safety buffer};
        \node [block, below of=safetybuffertext, text width=6.5em, node distance = 1.5cm, dashed] (restorea0) {
          \begin{tikzpicture}
            \node [text width=5.5em] (poptext) {pop \texttt{a0}};
            \node [text width=5.5em, draw, below of=poptext, node distance=0.8cm]   (innernode) {popped value};
          \end{tikzpicture}
        };
      \end{tikzpicture}
    };
    \node [block, below of=safetybuffer, node distance=2.2cm] (jalr) {\texttt{jalr a5}};
    \node [right of=branch, text width=6em, node distance=3cm] (writetarget) {write targets};
    \node [left of=safetybuffer, text width=6em, node distance=3cm](branchtarget) {branch target};
    \path [line] (begin) -- (popa5);
    \path [line] (popa5) -- (writea0);
    \path [line] (writea0) -- (writegadget);
    \path [line] (writegadget) -- (branch);
    \path [line] (safetybuffer) -- (jalr);
    \path [line] (-2.1, -6.2) -- (safetybuffer);
    \draw (branch) -| (branchtarget);
    \draw (writea0) -| (writetarget);
    \draw (writegadget) -| (writetarget);
    \path [line] (writetarget) |- (1.27, -6.2);
    \path [line] (writetarget) |- (1.1, -7.1);
  \end{tikzpicture}
  \caption{An example of using self-modifying ROP to perform a function call with an argument. Here, the value of \texttt{a0} must be restored prior to the function call since the branch gadget cobbles \texttt{a0}. In addition, a safety buffer is provided so that the called function's own stack allocations does not overwrite gadgets. }
  \label{figure:functioncall}
\end{figure}

\section{Unintended instructions as a result of the C extension}
As mentioned in \autoref{section:platform}, the C extension for RISC-V allows some level of variable-length instructions, meaning that unintended instructions are possible. For reasons explained in \autoref{section:nopgadget}, it is very likely that the unintended instruction sequence will later resync with intended program control flow to lead to a valid function epilogue. Out of 2837 gadgets we found in the version of glibc analyzed, we found a total of 711 gadgets that start at an entry point not found in the libc disassembly. 

Looking through the starting instructions of these gadgets, there are several that write to the \texttt{zero} register, several that modify the \texttt{ra} register before it is popped, and several that modify the temporary registers. We mostly ignore the temporary registers in our ROP gadgets because there are very few POP gadgets for them since they are caller saved. We also found lots of branches, nops, and floating point instructions. 

Some instructions that seemed useful were move, load and store instructions that used a wide variety of registers, as well as some instructions that moved the stack pointer (plus an offset) to several registers including \texttt{a0}. The latter could be used to perform self-modifying ROP in situations where the location of the stack pointer is not known, by performing arithmetic on the moved \texttt{sp} value and writing to the resulting address. 

Neither our example ROP gadgets nor the gadgets used in our ROP chain generator use unintended instructions, so our ROP chains and techniques should generalize to a RISC-V chip without the C extension.

\section{Implementing the Brainfuck instruction set in RISC-V ROP}
\label{section:brainfuckinstructions}
We demonstrate the Turing completeness of our found RISC-V ROP gadgets by creating a tool that, given an arbitrarily complex Brainfuck program, generates an equivalent RISC-V ROP chain. With some simplification, the Brainfuck instruction set is implemented as follows. The actual implementation takes care of many more preconditions/side effects, and also performs extra arithmetic since our readMEM/writeMEM gadgets are usually offsetted.

\begin{itemize}
  \item \textbf{Initialization}: Set \texttt{a0} to point the middle of a large buffer. This buffer will be the Brainfuck "tape". For simplicity, we will use a cell size of 64 bits rather than the standard 8 bits. We are free to make this modification since the Turing completeness of Brainfuck does not depend on integer overflow. \cite{davis_1966} \texttt{a0} will be the only register which is preserved across instructions other than \textbf{<} and \textbf{>}, and will represent the Brainfuck "pointer".
  \item \textbf{The > instruction}: Increment \texttt{a0} 8 times.
  \item \textbf{The < instruction}: Decrement \texttt{a0} 8 times.
  \item \textbf{The + instruction}: Write \texttt{a0} to future self-modified gadgets. Read the value of \texttt{(a0)} into \texttt{a0}. Increment \texttt{a0}. Using self-modifying ROP, pop the original address stored in \texttt{a0} into \texttt{s0} and write \texttt{a0} to \texttt{(s0)}. Finally, use self-modifying ROP again to restore the original value of \texttt{a0}. 
  \item \textbf{The - instruction}: Identical to above.
  \item \textbf{The . instruction}: Write \texttt{a0} to a future self-modified gadget. Read the value of \texttt{(a0)} into \texttt{a0}. Call \texttt{putchar}. Using self-modifying ROP, restore the original value of \texttt{a0}.
  \item \textbf{The , instruction}: Write \texttt{a0} to future self-modified gadgets. Call \texttt{getchar} to replace \texttt{a0} with user input. Using self-modifying ROP, pop the original address stored in \texttt{a0} into \texttt{s0} and write \texttt{a0} to \texttt{(s0)}. Use self-modifying ROP again to restore the original value of \texttt{a0}. 
  \item \textbf{The [ instruction}: Write \texttt{a0} to a self-modified \texttt{a0} pop gadget which immediately follows. (This is to restore the value after the unconditional branch in the \texttt{]} instruction.) Write \texttt{a0} to future self-modified gadgets. Dependent on \texttt{(a0) == 0}, perform a conditional branch either to the next instruction or to the instruction after the corresponding \texttt{]} instruction. In either case, the next gadget being executed should restore the value of \texttt{a0} through self-modifying ROP.
  \item \textbf{The ] instruction} Write \texttt{a0} to the self-modified gadget in the implementation of the corresponding \texttt{[} instruction. Perform an unconditional branch to the corresponding \texttt{[} instruction, so that it can restore the original value of \texttt{a0}.
  \item \textbf{Ending} Set \texttt{a0} to 0 and call \texttt{exit}.
\end{itemize}

\subsection{Compiling Brainfuck to RISC-V ROP}
\label{section:brainfucktool}
We introduce a compiler that compiles Brainfuck programs of arbitrary complexity into RISC-V ROP chains using the above mapping, written in pure JavaScript and designed to run in the browser. An accompanying C/assembly program reads the output of the browser tool from a file and populates memory using the output, then begins ROP execution by moving \texttt{sp} to a populated region and simulating a \texttt{NOP} gadget. 

\subsection{Results}
\label{section:brainfuckresults}
After trying several Brainfuck programs, with the caveat that Brainfuck programs that assume an 8-bit cell size will not work, the vast majority of Brainfuck programs converted to RISC-V ROP and ran easily with no issue, including an implementation of bubble sort (\href{https://raw.githubusercontent.com/adaptable-zz/bfbsort/master/sort.b}{link}), a program that prints arbitrarily many square numbers (\href{http://www.hevanet.com/cristofd/brainfuck/squares2.b}{link}), and a program that outputs an ASCII-art of the Sierpinski triangle (\href{http://www.hevanet.com/cristofd/brainfuck/sierpinski.b}{link}). 

\section{Conclusion and future work}
\label{section:conclusion}
We show that return-oriented programming on the RISCV64 architecture is fairly powerful, allowing for Turing-complete code execution on gadgets found in a version of GNU libc for Fedora Linux. We show that although difficult, it is possible to perform arbitrary calculation and to perform actions such as conditional branching, arithmetic, reading and writing memory, and calling libc functions with arguments. We introduce the technique of "self-modifying ROP chains" to solve several issues such as not having gadgets that move a value from one register to another. In order to demonstrate the power of these techniques, we created a compiler that converts Brainfuck code of arbitrary complexity into RISC-V ROP chains. 

In the future, we hope that more libraries for RISC-V, including other versions of libc for other operating systems, will be analyzed for ROP potential, and perhaps a more general and automated approach to analyzing and chaining ROP gadgets will make ROP attacks easier to carry out. 

%-------------------------------------------------------------------------------
\section*{Availability}
%-------------------------------------------------------------------------------

We make the Brainfuck-to-RISC-V-ROP compiler available at \href{https://garrettgu10.github.io/fuck-riscv-rop/}{https://garrettgu10.github.io/fuck-riscv-rop/} and we make its full source code available at \href{https://github.com/garrettgu10/fuck-riscv-rop}{https://github.com/garrettgu10/fuck-riscv-rop}. 

%-------------------------------------------------------------------------------
\bibliographystyle{plain}
\bibliography{\jobname}

%%%%%%%%%%%%%%%%%%%%%%%%%%%%%%%%%%%%%%%%%%%%%%%%%%%%%%%%%%%%%%%%%%%%%%%%%%%%%%%%
\end{document}